\documentclass[12pt]{iopart}
\usepackage{graphicx}

\begin{document}

\title{Proton capture cross section of $^{106,108}$Cd for the astrophysical p-process}
\author{Gy. Gy\"urky$^1$, G.G. Kiss$^1$, Z. Elekes$^1$, Zs. F\"ul\"op$^1$, E. Somorjai$^1$, T. Rauscher$^2$}
\address{$^1$ Institute of Nuclear Research (ATOMKI), P.O.Box 51, H-4001 Debrecen, Hungary}
\address{$^2$ Universit\"at Basel, CH-4056 Basel, Switzerland}
\ead{gyurky@atomki.hu}
\begin{abstract}
The proton capture cross sections of the two most proton rich, stable isotopes of Cadmium have been measured for the first time in the energy range relevant to the astrophysical p--process between E$_p$=\,2.4 and 4.7\,MeV. The $^{106}$Cd(p,$\gamma$)$^{107}$In and $^{108}$Cd(p,$\gamma$)$^{109}$In cross sections have been determined using the activation technique. Highly enriched as well as natural Cd targets have been irradiated with proton beams from both the Van de Graaff and Cyclotron accelerators of the ATOMKI. The cross sections have been derived by measuring the $\gamma$-radiation following the $\beta$-decay of the $^{107}$In and $^{109}$In reaction products. 
The experimental results are presented here and are compared with the predictions of Hauser-Feshbach statistical model calculations using the NON-SMOKER code. It is found that the calculation reproduces well the experimental data. The sensitivity of the model calculations to the proton- and $\gamma$-strengths is examined.
\end{abstract}
\submitto{\JPG}
\maketitle

\section{Introduction}
\label{sec:intro}
Despite the tremendous experimental and theoretical efforts of recent years,
the synthesis of the so called p--nuclei (the heavy, proton-rich isotopes which cannot be synthesized by neutron capture reactions in the s- or r-process) is still one of the least known processes of nucleosynthesis. It is generally accepted that the synthesis of the p--nuclei, the astrophysical p--process, involves mainly $\gamma$-induced reactions on abundant seed nuclei produced at earlier stages of nucleosynthesis by the s- (or to less extent the r-) process \cite{ar03}. During the p-process flow, material from the bottom of the valley of stability is driven to the proton rich side by subsequent ($\gamma$,n) reactions.
As the neutron separation energy increases while the charged-particle
separation energies decrease along this path, charged-particle
emitting ($\gamma,\alpha$) and ($\gamma$,p) reactions become increasingly
important for more proton-rich nuclei \cite{woo,rau06,rap06}. They can modify significantly
the final abundance distribution of p-nuclei by feeding isotopic chains
with lower charge number $Z$.

High energy photons with sufficient energy for dominating photodisintegration
appear
only in explosive nucleosynthetic scenarios. The generally accepted models
locate the p-process in the deep ONe-rich layers of massive stars either in their pre-supernova or supernova phases where temperatures of a few times 10$^9$~K are  reached.

Comprehensive modeling of the p-process requires on one hand detailed information on the stellar environment (temperature, original seed abundances,
burning time scale, etc.),
on the other hand nuclear physics also plays an important role.
In p-process modeling the reaction rates of hundreds of nuclear reactions involved in nuclear reaction networks must be known. All the reaction rates including the dominant $\gamma$-induced reactions are generally calculated with Hauser-Feshbach type statistical models.
The rates of $\gamma$-induced reactions can be calculated from the inverse capture reactions using the detailed balance theorem if the cross sections of the capture reactions are known experimentally.
While there are compilations for the (n,$\gamma$) cross sections,
very few charged particle induced reactions above the iron region have
been investigated experimentally, leaving the statistical model calculations largely untested. 

Realizing the need for testing experimentally the statistical model calculations in the region of proton rich nuclei, several (p,$\gamma$) and a few ($\alpha,\gamma$) reaction cross sections have been measured in recent years
and the results have been compared with model predictions (see e.g. \cite{gyu03} or \cite{har05} and references therein, and the KADoNiS database \cite{dil06}).
In general, the models are able to reproduce the experimental results within about a factor of two, however, some larger deviations are also found.
The existing experimental database is still not sufficient
to check the reliability of model calculations globally, therefore further experimental data are highly needed.

Recently, the ($\alpha,\gamma$) as well as the ($\alpha$,n) and ($\alpha$,p)
cross sections of $^{106}$Cd have been measured in the energy range relevant to the p-process and compared to statistical model calculations \cite{gyu06}.
An experiment is also in progress to obtain more information on the
important $\alpha$+nucleus optical potential via ($\alpha$,$\alpha$)
scattering on $^{106}$Cd; preliminary results are already available \cite{ki05}.
To give a complete description of $^{106}$Cd from the astrophysical
 p-process point of view, the measurement of its proton capture cross section
is also necessary.

In the present work the proton capture cross sections of $^{106}$Cd and $^{108}$Cd (the other p-isotope of Cadmium) have been measured.
A temperature window of $(2.5-3.5)\times 10^9$\,K for the astrophysical
reaction rates can be covered by our measured energy range. This
accommodates well the temperatures relevant for the p-process.
Consequently, the results can be compared with model predictions
right at the astrophysically relevant energies, no extrapolations are necessary.

\section{Investigated reactions}
\label{sec:invesreac}

Cadmium has 8 stable isotopes with mass numbers 106, 108, 110, 111, 112, 113, 114 and 116. The two lightest isotopes $^{106}$Cd and $^{108}$Cd are p-isotopes with low natural abundances of 1.25\% and 0.89\%, respectively.
Proton-capture on these two isotopes leads to unstable In isotopes ($^{107}$In and $^{109}$In) which decay by $\beta^+$ or electron capture to $^{107}$Cd and $^{109}$Cd, respectively. For both isotopes the $\beta$-decay is followed by $\gamma$-radiation which makes it possible to determine the proton-capture cross section using the activation method with $\gamma$-detection. In this method Cd targets are irradiated by a proton beam and the capture cross section is derived from the off-line measurement of the decay of reaction products.

Table 1 shows the decay parameters of the two residual In isotopes. Note that only the strongest $\gamma$-radiations following the $\beta$-decay of the reaction products are listed. 

\begin{table}
\caption{\label{tab:decay}Decay parameters of $^{107}$In and $^{109}$In isotopes. Only the strongest $\gamma$-radiations following the $\beta$-decay of the reaction products which were used for the analysis are shown. The data for $^{107}$In are taken from \cite{NDS1} and for $^{109}$In from \cite{NDS2} with the exception of the $^{109}$In half-life which is taken from a more recent work \cite{gyu05}.}
\lineup
\begin{indented}
\item[]\begin{tabular}{cccc}
\br
Product nucleus &
Half life &
Gamma energy [keV] &
Relative intensity \\ 
& [hour] & & per decay [\%]\\ 
\mr
$^{107}$In &  0.540 $\pm$ 0.005 & 204.96 & 47.2 $\pm$ 0.3 \\
$^{109}$In & 4.168 $\pm$ 0.018 & 203.50  & 73.5 $\pm$ 0.5 \\
\br
\end{tabular}
\end{indented}
\end{table}

Owing to the different decay patterns of the two reaction products,
it is possible to measure both cross sections in a single activation
experiment provided
the target contains both $^{106}$Cd and $^{108}$Cd isotopes.
Cadmium targets of natural isotopic abundance could in principle be
appropriate for the cross section determination.
Proton induced reactions on the heavier Cd isotopes, however, can be disturbing if they also lead to off-line $\gamma$-radiation.
In the astrophysically relevant low energy region the cross sections of the two investigated capture reactions are very low, thus the elimination of any disturbing $\gamma$-radiation from the spectra is highly needed.
Such disturbing $\gamma$-radiation can come e.g. from the
$^{110}$Cd(p,$\gamma$)$^{111}$In or $^{113}$Cd(p,n)$^{113}$In reactions, etc.
Therefore, in order to avoid the disturbing activity produced by heavier Cd isotopes, enriched targets were used in the lower energy region where the cross section and consequently the yield of the studied reactions is very low. 

Both produced In isotopes have metastable states. The proton capture on Cd can populate the ground as well as the metastable states in In. In the case of both isotopes, however, the metastable states decay exclusively 
by internal transition to the ground state. Moreover, the half-lives of the metastable states are short (54.6\,s for $^{107m}$In \cite{NDS1} and 1.34\,m for $^{109m}$In \cite{NDS2}). Therefore, after a suitable cooling time between the end of the irradiation and the beginning of the $\gamma$-counting, the metastable states have decayed completely to the grounds states and hence the measurement of the ground state decay provides information about the total capture cross section.

\section{Experimental procedure}
\label{sec:exp}
%\subsection{Target preparation}
%\label{subsec:target}

The experimental setup was similar to the one used for our previous Se(p,$\gamma$) experiment \cite{gyu03},
and described in details in a preliminary report on Cd(p,$\gamma$) \cite{gyu05a}. Shortly, thin Cd targets have been irradiated with proton beams from both the Van de Graaff and Cyclotron accelerators of the ATOMKI. The targets were prepared by evaporating natural or highly enriched (enrichment in $^{106}$Cd is 96.47\,\%) metallic Cd onto thin Al backing. The thickness of the targets has been measured by weighing. The energy range from E$_p$=\,2.4 to 4.7\,MeV was covered in 200\,keV steps using the Van de Graaff accelerator in the lower and the Cyclotron in the upper part of the energy range. Each irradiation lasted about 10 hours with a beam current of 500\,nA and the target stability was monitored by detecting the backscattered protons from the Cd target.

Table \ref{tab:irrad} shows the summary of the irradiations and the list of used targets. Note that the targets enriched in $^{106}$Cd contained 2.05\% $^{108}$Cd, still higher than the natural abundance of $^{108}$Cd (0.89\%). 

\begin{table}
\caption{\label{tab:irrad} Details of the irradiations and the targets used. The energy loss has been calculated with the SRIM code \cite{SRIM}}
\lineup
\begin{indented}
\item[]\hspace{-2cm}\begin{tabular}{cccr@{\hspace{1mm}$\pm$\hspace{1mm}}lc}
\br
Beam energy & accelerator & target type & \multicolumn{2}{c}{target thickness} & energy loss in target\\
\,[keV] & & & \multicolumn{2}{c}{[$\mu$g/cm$^2$]} & [keV]\\
\mr
2.400 & Van de Graaff	& enriched	& \hspace{5mm}595	& 42	& 36.6 \\
2.600 & Van de Graaff	& enriched	& 100	& 7	& 5.9 \\
2.800 & Van de Graaff	& enriched	& 165	& 12	& 9.3 \\
2.952 & cyclotron	& enriched	& 436	& 31	& 23.7 \\
3.000 & Van de Graaff	& natural	& 160	& 11	& 8.1 \\
3.200 & Van de Graaff	& natural	& 178 & 12	& 8.7 \\
3.200 & Van de Graaff	& enriched	& 100	& 7	& 5.2 \\
3.400 & Van de Graaff	& natural	& 160	& 11	& 7.5 \\
3.545 & cyclotron	& enriched	& 436	& 31	& 21.2 \\
3.600 & Van de Graaff	& natural	& 178	& 12	& 8.1 \\
3.744 & cyclotron	& enriched	& 146	& 10	& 6.8 \\
3.942 & cyclotron	& natural	& 290	& 20	& 12.5 \\
4.140 & cyclotron	& natural	& 246	& 17	& 10.3 \\
4.338 & cyclotron	& natural	& 246	& 17	& 10.0 \\
4.536 & cyclotron	& natural	& 290	& 20	& 11.4 \\
4.723 & cyclotron	& natural	& 246	& 17	& 9.4 \\
4.723 & cyclotron	& enriched	& 146	& 10	& 5.9 \\

\br
\end{tabular}
\end{indented}
\end{table}

\begin{figure}
\begin{indented}
\item[]\resizebox{0.8\textwidth}{!}{\rotatebox{270}{\includegraphics{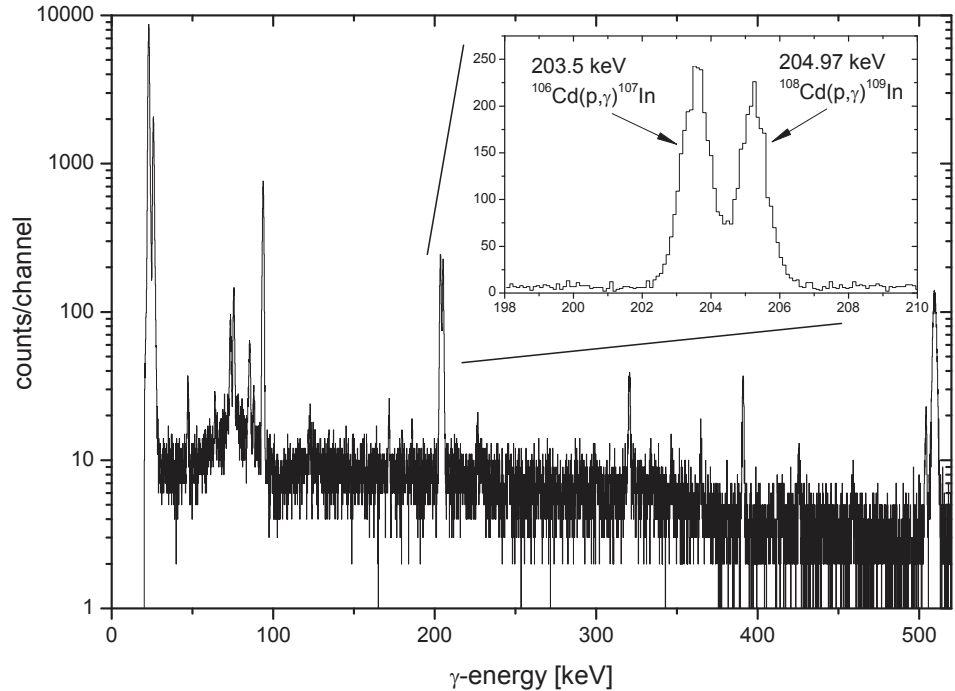}}}
\caption{Typical activation $\gamma$-spectrum taken after the irradiation of a natural 
Cd target with a 3.8 MeV proton beam. This spectrum was taken for 3 hours 
starting 8 minutes after the end of the irradiation. The inset shows the two resolved 
$\gamma$-peaks from the two reactions studied. The higher energy peaks visible in the spectrum are coming from beam-induced activities on heavier Cd isotopes.}
\label{fig:spectrum}
\end{indented}
\end{figure}

The induced $\gamma$-activity of the targets has been measured with a
calibrated HPGe detector equipped with 5\,cm thick lead shielding. The $\gamma$-spectra were taken for at least 10
hours to follow the decay of both produced In isotopes.
The strongest $\gamma$-radiations from the decay of the two In isotopes are
very close to each other in energy (203.5 and 204.97\,keV). However, the
energy resolution of the HPGe detector at this low energy is about 0.8\,keV
(FWHM), hence the two peaks could be resolved (see the inset of
Fig.\,\ref{fig:spectrum}). Moreover, the different half-lives of the two
reaction products makes the separation even easier. At the beginning of the counting period the 204.96\,keV peak from $^{107}$In decay dominates, while towards the end of the 10 hour's counting time this isotope has already decayed and the 203.5\,keV peak form $^{107}$In decay remains. The inset in Fig.\,\ref{fig:spectrum} shows an example where the two peak amplitudes are similar.

\section{Results and conclusion}
\label{sec:results}

The measured cross sections of both investigated reactions
cover more than 3 orders of magnitude from 3 to 5000\,$\mu$b.
Table~\ref{tab:res} lists the experimental results for both reactions in the
form of cross sections as well as astrophysical S-factors.
The rows in Table~\ref{tab:res} correspond directly to the rows of
Table~\ref{tab:irrad}. There is good agreement between points measured at
similar energies but with different targets (natural or enriched) or
different accelerators.

The quoted uncertainty in the E$_{c.m.}$ values correspond to the energy stability of the proton beam and to the uncertainty of the energy loss in the target. The uncertainty of the cross section values is the quadratic sum of the following partial errors: efficiency of the HPGe detector (7\%), number of target atoms (6\%), current measurement (3\%), and counting
statistics (up to 12\%).
\begin{table}
\caption{\label{tab:res} Experimental cross section and S factor of the $^{106}$Cd(p,$\gamma$)$^{107}$In and $^{108}$Cd(p,$\gamma$)$^{109}$In reactions}
\lineup
\begin{indented}
\item[]\hspace{-2cm}\begin{tabular}{r@{\hspace{1mm}$\pm$\hspace{1mm}}lr@{\hspace{1mm}$\pm$\hspace{1mm}}lr@{\hspace{1mm}$\pm$\hspace{1mm}}lr@{\hspace{1mm}$\pm$\hspace{1mm}}lr@{\hspace{1mm}$\pm$\hspace{1mm}}lr@{\hspace{1mm}$\pm$\hspace{1mm}}l}
\br
\multicolumn{6}{c}{$^{106}$Cd(p,$\gamma$)$^{107}$In} & \multicolumn{6}{c}{$^{108}$Cd(p,$\gamma$)$^{109}$In} \\
\mr
\multicolumn{2}{c}{E$_{c.m.}$} & \multicolumn{2}{c}{cross section} & \multicolumn{2}{c}{S factor} & \multicolumn{2}{c}{E$_{c.m.}$} & \multicolumn{2}{c}{cross section} & \multicolumn{2}{c}{S factor} \\ 
\multicolumn{2}{c}{[MeV]} & \multicolumn{2}{c}{[$\mu$b]} &  \multicolumn{2}{c}{[10$^6$ MeV$\cdot$b]} & \multicolumn{2}{c}{[MeV]} & \multicolumn{2}{c}{[$\mu$b]} &  \multicolumn{2}{c}{[10$^6$ MeV$\cdot$b]}\\ 
\mr
2.359	& 0.003	& 3.71	& \0\00.43	& 201\0	& 23 & 2.360	&	0.003	&	3.46	&	\0\00.45	&	187\0	&	24\\
2.573	& 0.003	& 7.96	& \0\00.95	& 127\0	& 15 & 2.573	&	0.003	&	11.6	&	\0\02.0	&	186\0	&	32\\
2.769	& 0.003	& 21.2\0	& \0\02.5	& 126\0	& 15 & 2.770	&	0.003	&	27.3	&	\0\03.3	&	162\0	&	20	\\
2.913	& 0.009	& 52.9\0	& \0\06.1	& 162\0	& 19 & 2.913	&	0.009	&	70.0	&	\0\08.1	&	216\0	&	25\\
2.968	&	0.003	&	\multicolumn{4}{c}{yield too small for analysis} & 2.968	&	0.003	&	58.0 &	\0\07.4	&	141\0	&	18	\\
3.166	& 0.003	& 71.5\0	& \0\09.5	& 77.2	& 10.3 & 3.166	&	0.003	&	157\0	&	\019	&	170\0	&	21	\\
3.168	& 0.003	& 63.3\0	& \0\07.3	& 67.8	& \07.8 & 3.168	&	0.003	&	146\0	&	\017	&	157\0	&	18 \\
3.364	& 0.003	& 209\0\0	& \026	& 108\0 & 13 & 3.365	&	0.003	&	298\0	&	\035	&	154\0	&	18	\\
3.501	& 0.011	& 223\0\0	& \026	& 72.3	& \08.4 & 3.505	&	0.011	&	482\0	&	\057	&	155\0	&	18	\\
3.562	& 0.004	& 295\0\0	& \037	& 78.4	& \09.8 & 3.563	&	0.004	&	391\0	&	\045	&	104\0	&	12	\\
3.706	& 0.011	& 346\0\0	& \040	& 58.5	& \06.8 & 3.706	&	0.011	&	686\0	&	\079	&	116\0	&	13	\\
3.899	& 0.012	& 528\0\0	& \064	& 50.8	& \06.2 & 3.900	&	0.012	&	825\0	&	\097	&	79.3	&	\09.3	\\
4.096	& 0.012	& 936\0\0	& 114	& 52.8	& \06.4 & 4.097	&	0.012	&	1862\0	&	219	&	105\0	&	12	\\
4.293	& 0.013	& 1232\0\0	& 149	& 42.4	& \05.1 & 4.293	&	0.013	&	2744\0	&	322	&	94.6	&	11	\\
4.488	& 0.014	& 1283\0\0	& 155	& 27.9	& \03.4 & 4.489	&	0.014	&	2592\0	&	304	&	56.5	&	\06.6	\\
4.674	& 0.014	& 2329\0\0	& 280	& 33.8	& \04.1 & 4.675	&	0.014	&	5392\0	&	633	&	78.1	&	\09.2	\\
4.676	& 0.014	& 1918\0\0	& 221	& 27.7	& \03.2 & 4.677	&	0.014	&	4733\0	&	545	&	68.3	&	\07.9	\\

\br
\end{tabular}
\end{indented}
\end{table}

\begin{figure}
\resizebox{\columnwidth}{!}{\rotatebox{270}{\includegraphics{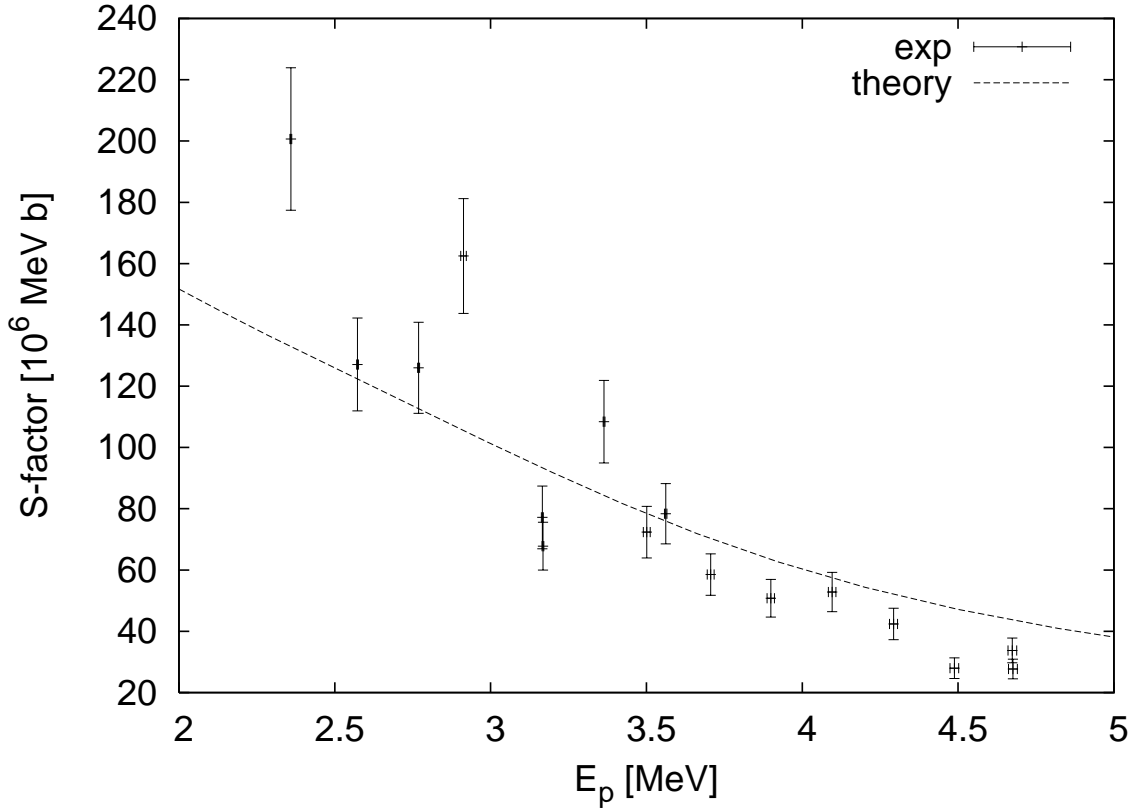}}}
\caption{Astrophysical S-factor of the $^{106}$Cd(p,$\gamma$)$^{107}$In
reaction as a function of the proton c.m. energy. The dashed line is the 
global Hauser-Feshbach
statistical model prediction of the NON-SMOKER code \cite{NON-SMOKER}.}
\label{fig:res106}
\end{figure}

\begin{figure}
\resizebox{\columnwidth}{!}{\rotatebox{270}{\includegraphics{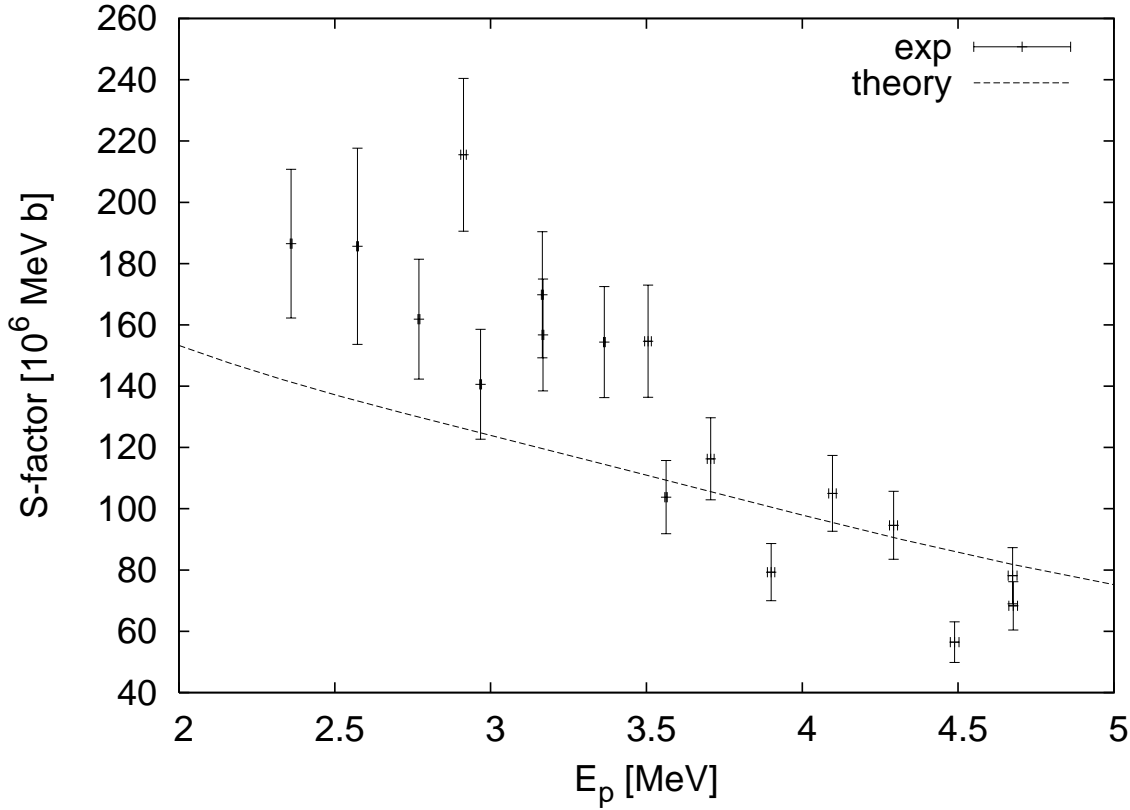}}}
\caption{Astrophysical S-factor of the $^{108}$Cd(p,$\gamma$)$^{109}$In
reaction as a function of the proton c.m. energy. The dashed line is the 
global Hauser-Feshbach
statistical model prediction of the NON-SMOKER code \cite{NON-SMOKER}.}
\label{fig:res108}
\end{figure}

The astrophysical S-factors are plotted in Figs.\ \ref{fig:res106} and
\ref{fig:res108}. They are compared to the results of a Hauser-Feshbach
statistical model calculation with the NON-SMOKER code \cite{rath,NON-SMOKER}.
It should be noted that the standard calculation is made with global
parameters which are not adjusted to any local nuclear or reaction
properties. There is good agreement between experiment and theory across
the measured energy range for the reaction $^{106}$Cd(p,$\gamma$)$^{107}$In.
There is also good agreement in the case of $^{108}$Cd(p,$\gamma$)$^{109}$In
for energies above 3.5 MeV whereas there seems to be a slight systematic
underprediction below that energy. However, it is still well within the
usual 30\% error which has to be assigned to the global model \cite{rtk97,dr06}.
Although it appears as if the data are systematically higher below 3.5 MeV
in the case of $^{108}$Cd, no correlation with either the used targets or
the accelerator type was found in a closer inspection of Table \ref{tab:irrad}.

\begin{figure}
\resizebox{\columnwidth}{!}{\rotatebox{270}{\includegraphics{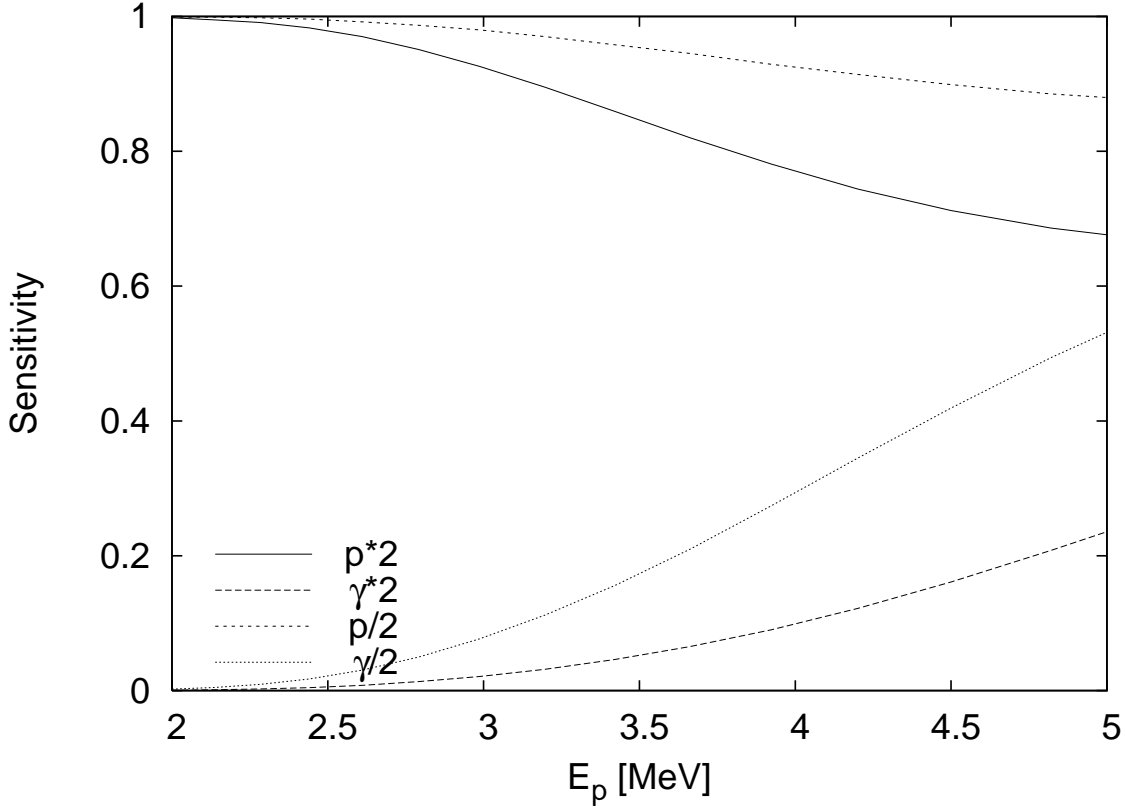}}}
\caption{Sensitivity of the $^{106}$Cd(p,$\gamma$)$^{107}$In
cross sections to a variation in the proton-
and $\gamma$-strengths as a function of the proton c.m. energy.
The sensitivity ranges from 0 (no change) to 1 (the cross section is
changed by the same factor as the strength).}
\label{fig:sens106}
\end{figure}

\begin{figure}
\resizebox{\columnwidth}{!}{\rotatebox{270}{\includegraphics{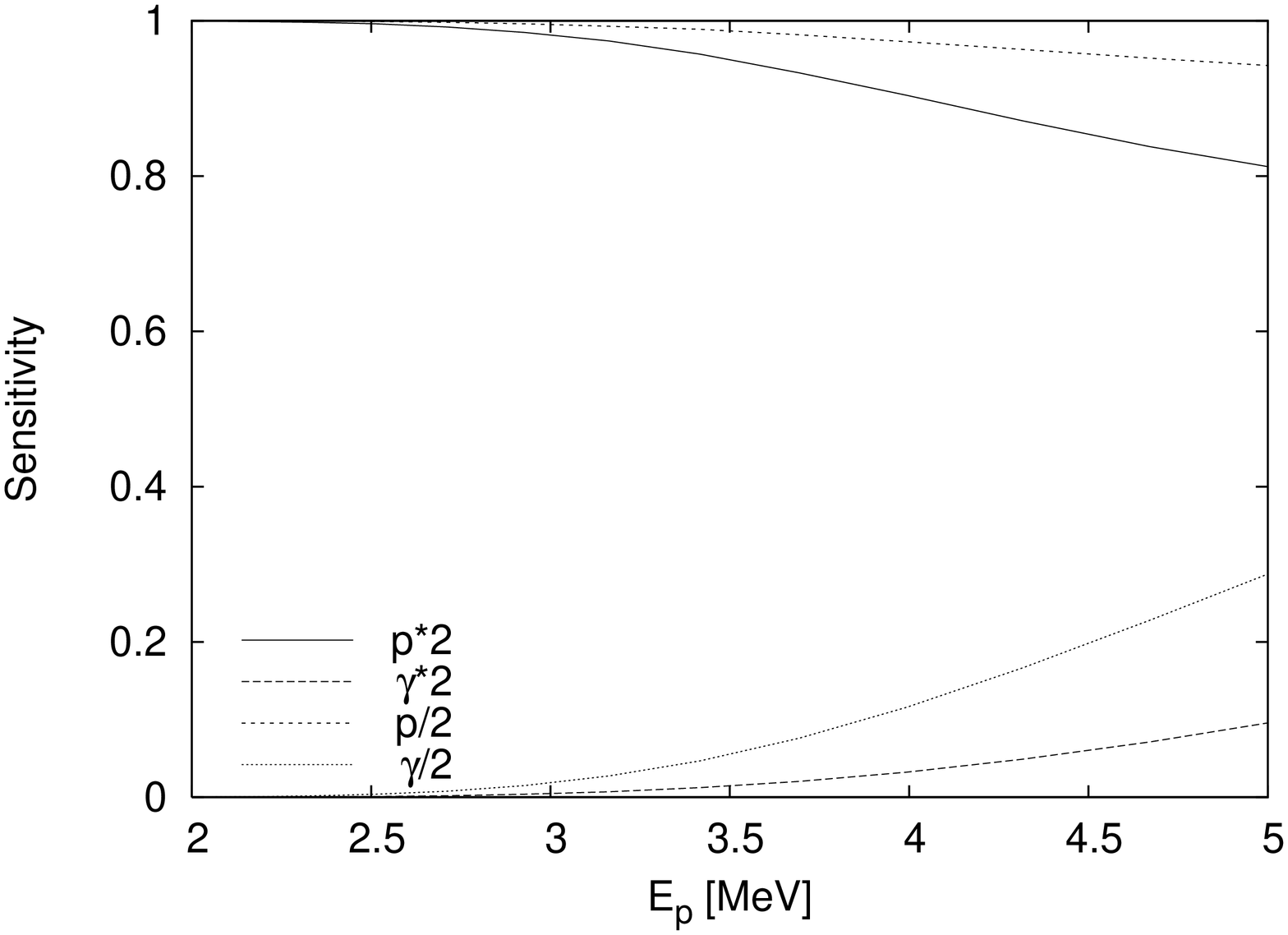}}}
\caption{Same as Fig.\ \ref{fig:sens106} but for the reaction
$^{108}$Cd(p,$\gamma$)$^{109}$In.}
\label{fig:sens108}
\end{figure}

From Figs.\ \ref{fig:res106} and \ref{fig:res108} one might imply that the
energy-dependence of the theoretical S-factor may be slightly too flat. A
stronger increase towards lower energies would accommodate the measurements
well for both reactions. The calculated cross sections and S-factors are
mainly dependent on the proton- and $\gamma$-strengths. In order to study
the sensitivity we have performed calculations to study the impact of a
variation of the strengths by a factor of two. The result is shown in Figs.\
\ref{fig:sens106} and \ref{fig:sens108}. The plotted sensitivity can
assume values between 0 and 1, ranging from no impact to a full factor of two
change in the cross section and S-factor. It can be seen that the
S-factors of both reactions are more sensitive to the proton strengths
than the $\gamma$-strengths across the given energy range. This may appear
surprising at first glance but it has to be realized that the proton
widths are smaller than the $\gamma$-widths due to the
Coulomb suppression at such low energies. Consequently, a change in the
nuclear properties determining the $\gamma$-strength would only have a
limited impact and only at the high end of the measured range. The impact
would be larger for $^{108}$Cd(p,$\gamma$)$^{109}$In than for the other
reaction.

The E1 GDR strength function and the nuclear level density in the compound
nucleus determine the $\gamma$-strength whereas the main property for the
proton width is the optical potential for protons. The standard
calculation shown here uses E1 and M1 transmission coefficients
as described in \cite{rath} (additionally, E2 transitions were considered
but they do not contribute significantly here), the
nuclear level density of \cite{rtk97}, and the microscopic proton
potential of \cite{jlm} with low-energy modifications as described in
\cite{rath}. However, since there
is reasonable reproduction of the data within the quoted errors it is
impossible to make a case for a required modification of any of these
properties.

The central quantity important for the application in astrophysical models
is the astrophysical reaction rate obtained by folding the cross sections
with the temperature dependent Maxwell-Boltzmann distribution of projectile
energies. Because of the limited energy range of the data, the relevant
integration can only be performed for a limited number of temperatures when
adopting an upper limit for the error in the integration due to the
energy cutoff. For the derivation of the astrophysical reaction rates
from the data, we used the code EXP2RATE \cite{exp2rate} which
automatically determines the possible temperature window and also accounts
for the experimental error bars. Allowing for an error of at most 15\%,
we find that the covered temperature range is $(2.5-3.75)\times 10^9$\,K
for $^{106}$Cd(p,$\gamma$)$^{107}$In 
and $(2.5-3.5)\times 10^9$\,K for $^{108}$Cd(p,$\gamma$)$^{109}$In. This
covers well the relevant p-process temperature range which is usually
given as $(2.0-3.0)\times 10^9$\,K \cite{ar03,rau06}. The
resulting rates with their experimental errors are given in Tables
\ref{tab:rates106} and \ref{tab:rates108}.

\begin{table}
\caption{\label{tab:rates106}Astrophysical reaction rates for
$^{106}$Cd(p,$\gamma$)$^{107}$In as
function of stellar temperature $T$;
only those temperatures are given for which the rates
can be computed from the data. Also included are the experimental errors.
For comparison, also the theoretical laboratory
rates of \cite{NON-SMOKER} are given,
along with the ratio of experimental rate to theoretical rate and the
stellar enhancement factors SEF.}
\lineup
\begin{indented}
\item[]\begin{tabular}{lr@{\hspace{1mm}$\pm$\hspace{1mm}}llr@{\hspace{1mm}$\pm$\hspace{1mm}}lr}
\br
\multicolumn{1}{c}{$T$}&\multicolumn{2}{c}{exp}&\multicolumn{1}{c}{theory}&
\multicolumn{2}{c}{exp/theory}&\multicolumn{1}{c}{SEF}\\
\multicolumn{1}{c}{[10$^9$ K]}&\multicolumn{2}{c}{[cm$^3$ s$^{-1}$ mole$^{-1}$]}&\multicolumn{1}{c}{[cm$^3$ s$^{-1}$ mole$^{-1}$]}&\multicolumn{3}{c}{ }\\
\mr
2.50&1.7797&0.2191&1.92&0.93&0.11&1.01 \\
2.75&5.6627&0.7026&\multicolumn{4}{c}{ } \\
3.00&(1.5186&$0.1893)\times 10^{1}$&$1.61\times 10^1$&0.94&0.12&1.03 \\
3.25&(3.5578&$0.4445)\times 10^{1}$&\multicolumn{4}{c}{ } \\
3.50&(7.4729&$0.9337)\times 10^{1}$&$8.50\times 10^1$&0.88&0.11&1.05 \\
3.75&(1.4346&$0.1790)\times 10^{2}$&\multicolumn{4}{c}{ } \\
\br
\end{tabular}
\end{indented}
\end{table}

\begin{table}
\caption{\label{tab:rates108}Same as Table \ref{tab:rates106} but for
the reaction
$^{108}$Cd(p,$\gamma$)$^{109}$In.}
\lineup
\begin{indented}
\item[]\begin{tabular}{lr@{\hspace{1mm}$\pm$\hspace{1mm}}llr@{\hspace{1mm}$\pm$\hspace{1mm}}lr}
\br
\multicolumn{1}{c}{$T$}&\multicolumn{2}{c}{exp}&\multicolumn{1}{c}{theory}&
\multicolumn{2}{c}{exp/theory}&\multicolumn{1}{c}{SEF}\\
\multicolumn{1}{c}{[10$^9$ K]}&\multicolumn{2}{c}{[cm$^3$ s$^{-1}$ mole$^{-1}$]}&\multicolumn{1}{c}{[cm$^3$ s$^{-1}$ mole$^{-1}$]}&\multicolumn{3}{c}{ }\\
\mr
2.50&2.5625&0.3391&2.38&1.08&0.14&1.00 \\
2.75&8.4221&1.1047&\multicolumn{4}{c}{ } \\
3.00&(2.3267&$0.3040)\times 10^{1}$&$2.17\times 10^1$&1.07&0.14&1.01 \\
3.25&(5.5990&$0.7313)\times 10^{1}$&\multicolumn{4}{c}{ } \\
3.50&(1.2044&$0.1577)\times 10^{2}$&$1.24\times 10^{2}$&0.97&0.13&1.00 \\
\br
\end{tabular}
\end{indented}
\end{table}

Comparison of the experimental with the theoretical rates shows that they
are similar, i.e.\ compatible with a ratio of unity,
within the experimental errors for all given temperatures and
for both reactions despite the small discrepancies between experiment and
prediction discussed above. This is due to the fact that the calculation
of the astrophysical reaction rate involves a weighted average across an
appropriate energy range and tends to cancel small deviations.

Finally, a further effect has to be considered under astrophysical conditions.
While the target nuclei are always in their ground states in a laboratory
experiment, nuclei in astrophysical plasmas can be thermally excited
according to the temperature. This has to be taken into account
in the calculation of the reaction rate.
Experimental and theoretical ground state rates are
compared in Tables \ref{tab:rates106} and \ref{tab:rates108}. However,
the effect of thermal excitation is negligible for the cases considered
here as is proven by the thermal enhancement factor SEF=$r_*/r_\mathrm{gs}$
which compares the ground state rate $r_\mathrm{gs}$ to the proper
stellar rate $r_*$ and is also shown in the two tables.

Summarizing, for the first time
we have performed measurements of the two reactions
$^{106}$Cd(p,$\gamma$)$^{107}$In and $^{108}$Cd(p,$\gamma$)$^{109}$In
in the energy range relevant for p-process nucleosynthesis. S-factors and
experimental reaction rates were derived. The results excellently confirm
the theoretical proton capture rates used in astrophysical models for
these reactions and thus help to narrow the search for the cause of
the remaining deficiencies
in the nuclear as well as astrophysical p-process modeling.

\ack{This work was supported by OTKA (T42733, T49245, F43408, D48283) and in
part by the Swiss National Science Foundation
(grants 200020-105328, 200020-113984). Zs. F\"ul\"op and Gy. Gy\"urky are Bolyai fellows.}

\end{document}